\documentclass[12pt,preprint]{aastex}

\shorttitle{The hot core of NGC1333-IRAS4A}
\shortauthors{Bottinelli et al.}

\newcommand{\kms}          {\mbox{${\rm km~s^{-1}}$}}
\newcommand{\pp}           {\noindent\hangindent 20pt\hangafter=1}

\begin{document}

\title{Complex molecules in the hot core of the low mass protostar 
NGC1333-IRAS4A}
\author{S. Bottinelli\altaffilmark{1,2}, C. Ceccarelli\altaffilmark{1}, 
B. Lefloch\altaffilmark{1}, J. P. Williams\altaffilmark{2}, 
A. Castets\altaffilmark{3}, E. Caux\altaffilmark{4},
S. Cazaux\altaffilmark{5},
S. Maret\altaffilmark{1}, B. Parise\altaffilmark{4}
,  A. G. G. M. Tielens\altaffilmark{6}}

\altaffiltext{1}{Laboratoire d'Astrophysique de l'Observatoire de Grenoble, 
BP 53, 38041 Grenoble, Cedex 9, France. \\
sbottine@obs.ujf-grenoble.fr; 
ceccarel@obs.ujf-grenoble.fr; 
lefloch@obs.ujf-grenoble.fr;
maret@obs.ujf-grenoble.fr}
\altaffiltext{2}{Institute for Astronomy, University of Hawai`i, 
2680 Woodlawn Drive, Honolulu HI 96822, USA.\\
jpw@ifa.hawaii.edu}
\altaffiltext{3}{Observatoire de Bordeaux, 2 Rue de l'Observatoire, BP 89,
33270 Floirac, France.\\
castets@obs.u-bordeaux1.fr}
\altaffiltext{4}{Centre d'Etude Spatiale des Rayonnements, CNRS-UPS, 9 Avenue
du Colonel Roche, BP 4346, 31028 Toulouse, Cedex 4, France.\\
caux@cesr.fr; parise@cesr.fr}
\altaffiltext{5}{INA Osservatorio Astrofisico d'Arcetri, 1 Ple Aldo Moro,
Florence, Italy. \\cazaux@arcetri.astro.it}
\altaffiltext{6}{Kapteyn Astronomical Institute, BO Box 800, 9700 
AV Groningen, The Netherlands. \\tielens@astro.rug.nl}

\begin{abstract}
We report the detection of complex molecules (HCOOCH$_3$, HCOOH and
CH$_3$CN), signposts of a ``hot core''
like region, toward the low mass, Class 0 source NGC1333-IRAS4A. 
This is the second low mass protostar where such complex molecules have been
searched for and reported, the other source being IRAS16293--2422.
It is therefore likely that compact (few tens of AUs) regions of 
dense and warm gas, where the
chemistry is dominated by the evaporation of grain mantles, and where
complex molecules are found, are common in low mass Class 0 sources.
Given that the chemical formation timescale is much shorter than
the gas hot core crossing time, it is not clear whether
the reported complex molecules are formed on the grain surfaces 
(first generation molecules) or in the warm gas by reactions involving
the evaporated mantle constituents (second generation molecules).
We do not find evidence for large differences in the 
molecular abundances, normalized to the formaldehyde abundance, between the 
two solar type protostars, suggesting perhaps a common origin.
\end{abstract}


\section{Introduction}

There is strong support --- from the composition of cometary and meteoritic
materials --- for the notion that the solar nebula, from which the planets
formed, passed through a phase of warm, dense gas with a rich chemistry. 
While much observational effort has been dedicated to the study of such
hot cores around massive protostars, hot cores around low
mass protostars have received little attention.  Only very recently has
the first hot core around a solar-type protostar been discovered towards
the typical Class 0 source, IRAS16293--2422 (hereafter IRAS16293), 
exhibiting all characteristics of such regions: 
warm temperatures ($>100$ K) and high densities 
($>10^7$ cm$^{-3}$: Ceccarelli et al. 2000a), 
high abundances of hydrides (CH$_3$OH, H$_2$CO, H$_2$O: 
Ceccarelli et al. 2000a, b; Sch\"oier et al. 2002), 
high deuteration levels ($>10$\%: Ceccarelli et al. 1998, 2001; Parise et al.
2002; Roberts et al. 2002), 
and complex molecules (HCOOCH$_3$, HCOOH, CH$_3$OCH$_3$, CH$_3$CN, 
C$_2$H$_5$CN: Cazaux et al. 2003). 
The definition of ``hot core'' used
for massive protostars implies the presence of a relatively large amount
of warm and dense gas, along with a complex chemistry
triggered by the grain mantle evaporation (e.g. Walmsley et al. 1992).
In order to make clear that hot cores of low and high mass protostars
are, however, substantially different in the involved amount of material, 
we will use hereinafter the term ``{\bf hot corino}'' to identify 
the warm inner regions of the envelope surrounding the low mass protostars. 

The chemical composition of the (massive) hot cores is thought to reflect a
variety of sequential processes
(Wamsley et al. 1992; Charnley et al. 1992; Caselli et al. 1993; 
Charnley 1995; Rodgers \& Charnley 2001, 2003). 
Specifically, in the pre-collapse cold cloud phase, simple molecules 
form on grain surfaces by hydrogenation of CO and other heavy elements
(notably examples are H$_2$CO, CH$_3$OH and H$_2$S).
Upon heating by a newly formed star, these molecules, called ``first
generation'' or ``parent'' molecules, evaporate into the gas and undergo fast
neutral-neutral and ion-neutral reactions producing complex organic
molecules, i.e. ``second generation'' or ``daughter'' molecules.  
The first part of
this sequence, i.e. the formation of fully hydrogenated molecules on
the grain surfaces, has been demonstrated to occur in low mass
protostars too, for example by studies of multiply deuterated
molecules: formaldehyde (Ceccarelli et al. 1998, Bacmann et al. 2003),
methanol (Parise et al. 2002, 2004) and sulfide (Vastel et al. 2003).
Evaporation from grain mantles of these first generation species
(in particular H$_2$CO and CH$_3$OH) has been observed in IRAS16293
(Ceccarelli et al. 2000b, Sch\"oier et al. 2002) and in about a dozen
low mass protostars (Maret et al. 2004).  However, since the timescale
necessary to convert first generation molecules into complex, second
generation molecules (around $10^4-10^5$ yr; e.g. Charnley et al.
1992, 2001) is much longer than the transit time of the gas in the hot
corinos (few 100 yr; e.g. Sch\"oier et al. 2002), the formation 
in the gas of second generation molecules seems improbable
(e.g. Sch\"oier et al. 2002). 
The detection of a high abundance of complex
molecules in the hot core of IRAS16293 (Cazaux et al.
2003) has evidently been a challenge to the simple theoretical
sequence described above.  
The key question has shifted from ``Is a hot core present in low mass
protostars?'' to ``What is the origin of molecular complexity 
in these sources?''
In particular, there may well be chemical pathways to complex molecules
involving grain surface networks (e.g. Charnley 1995).
In order to answer this question, more observations in
other low mass protostars are necessary. This will allow the development of a
solid observational framework within which we might search for clues to the
formation of second generation molecules.
As remarked in previous studies, the question is
far from being academic, since the molecules in the hot corinos constitute
the material which will eventually form the proto-planetary disk and,
possibly, the planets of the forming Sun-like star.

In this Letter we present the first results of a survey we are carrying out
on the sample of Class 0 sources studied by Maret et al. (2004).
Here we report the detection of complex, second generation molecules in 
NGC1333-IRAS4A (hereafter IRAS4A),
a well known Class 0 protostar, and a target of several studies of molecular 
emission (e.g. Blake et al. 1995).
IRAS4A is part of the binary system IRAS 4, located in the NGC1333 reflection
nebula, in the Perseus cloud. It is separated by $31''$
from the other component, IRAS 4B, and was itself resolved into two
components with a separation of $2''$, by Lay et al. (1995).
The distance to the NGC1333 cloud is uncertain (see e.g. Maret et al. 
2002), but assuming a value
of 220~pc (derived by \u{C}ernis 1990, for consistency with previous work),
IRAS4A has a luminosity of $6~L_\odot$ and an 
envelope mass of $3.5~M_\odot$ (Sandell et al. 1991).
IRAS4A is associated with a very highly collimated outflow, detected in 
CO, CS, and SiO (Blake et al. 1995, Lefloch et al. 1998). Infall motion was 
detected by Di Francesco et al. (2001) and Choi et al. (1999)
with an estimated accretion rate of $1.1\times10^{-4}~M_\odot$~yr$^{-1}$,
an inner mass of 0.7~$M_\odot$ and an age of $\sim$6500 yr (see also Maret 
et al. 2002).

\section{Observations and results}

The observations were carried out in June 2003 with the IRAM 30-meter 
telescope.
The position used for pointing was 
$\alpha$(2000) = $03^{\rm h}29^{\rm m}10\fs 3$ and 
$\delta$(2000) = $31^\circ 13'31''$. Based on the observations of IRAS16293
by Cazaux et al. (2003), we targeted the following complex molecules:
methyl formate, HCOOCH$_3$ (A and E), 
formic acid, HCOOH, dimethyl ether, CH$_3$OCH$_3$, methyl cyanide, CH$_3$CN, 
and ethyl cyanide, C$_2$H$_5$CN. Different telescope settings were used
in order to include as many transitions as possible for each molecule.
All lines were observed with a low resolution,
1~MHz filter bank of 4~$\times$~256 channels split between different receivers,
providing a velocity resolution of $\sim$ 3, 2, and 1~\kms 
~at 3, 2, and 1~mm,
respectively. Each receiver was simultaneously connected to a unit of the 
autocorrelator, with spectral resolutions of 20, 80 or 320 kHz and bandwidths
between 40 and 240 MHz, equivalent to a (unsmoothed) velocity resolution 
range of 0.1--0.4~\kms. 
Typical system temperatures were 100--200~K, 180--250~K 
and 500--1500~K, at 3, 2 and 1~mm respectively.

Two observation modes were used: position switching with the OFF position at
an offset of $\Delta \alpha$ = --100$''$, $\Delta \delta$ = +300$''$, and 
wobbler switching with a 110$''$ throw in azimuth. Pointing and focus were
regularly checked using planets or strong quasars, providing a pointing 
accuracy of 3$''$. All intensities reported in this paper 
are expressed in units of main-beam brightness temperature. 
At 3, 2 and 1~mm, the angular resolution is 24, 16 and 10$''$ and
the main beam efficiency is 76, 69 and 50\%, respectively.

\begin{figure}
\epsscale{0.8}
\plotone{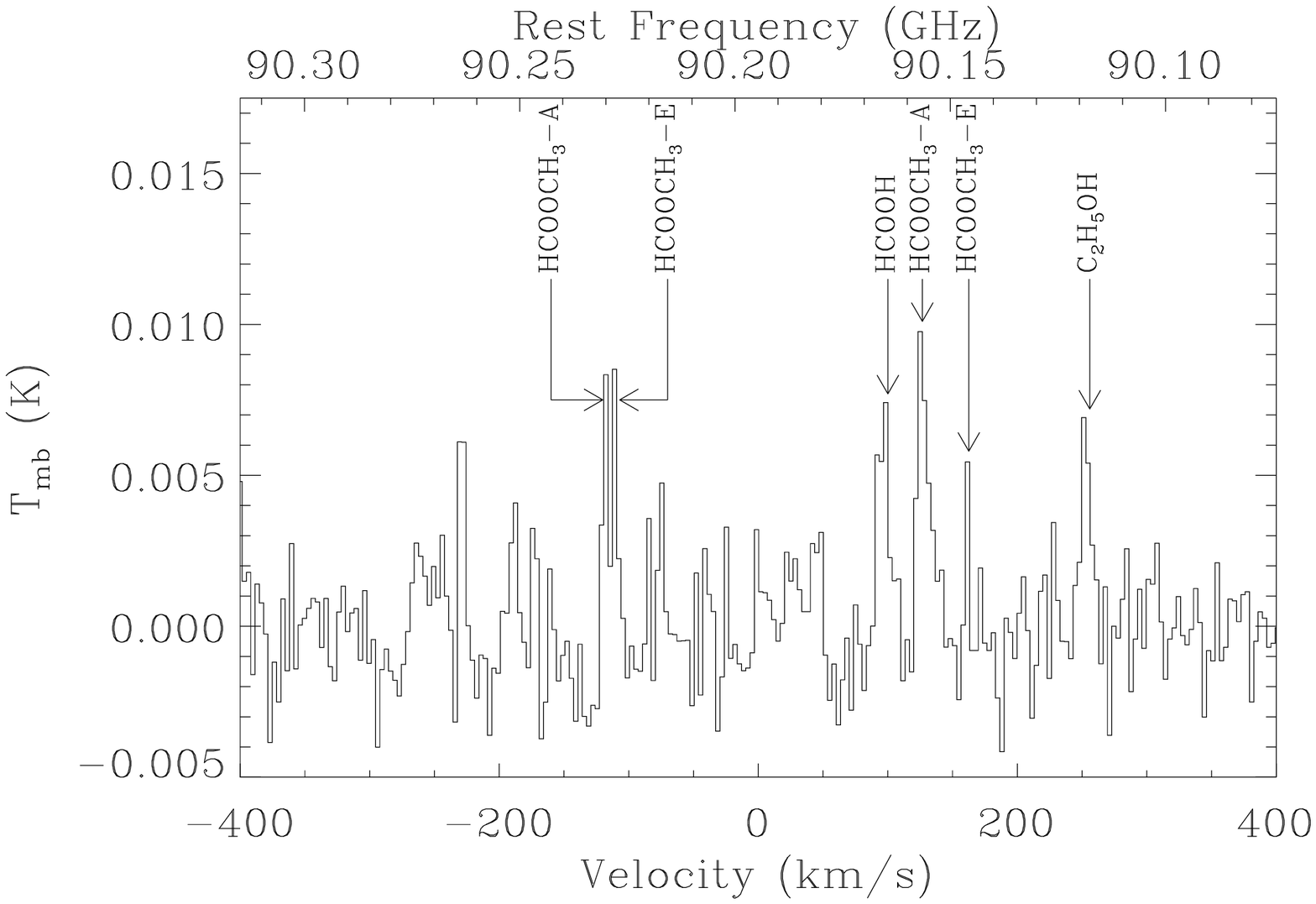}
\plotone{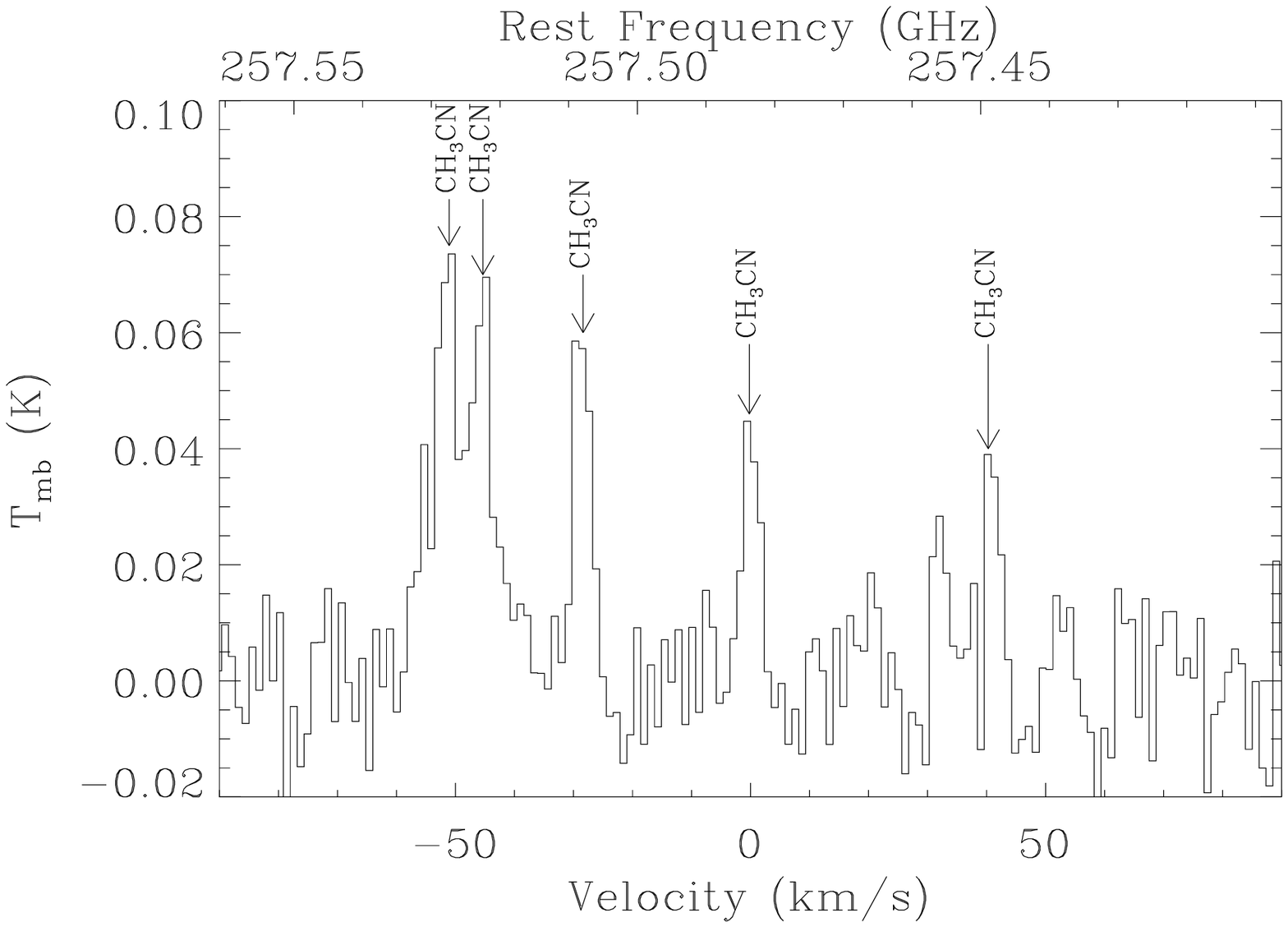}
\figcaption{Two low resolution spectra obtained during our observations of 
IRAS4A. Lines which are not labelled are unidentified. 
The rms noise level is 2~mK (top spectrum) and 12~mK (bottom spectrum).
The spectral resolution is 3.3 \kms (top) and 1.2 \kms (bottom).
The V$_{\rm LSR}$ is 7.0 \kms.
Known transitions are indicated but not all of them are detections, e.g.
HCOOCH$_3$ at 90.145 GHz is not considered as such, but the upper limit
derived from it is consistent with the rotational diagram of Fig.~\ref{rotdia}.
\label{spectrum}}
\end{figure}

\begin{figure}
\epsscale{0.8}
\plotone{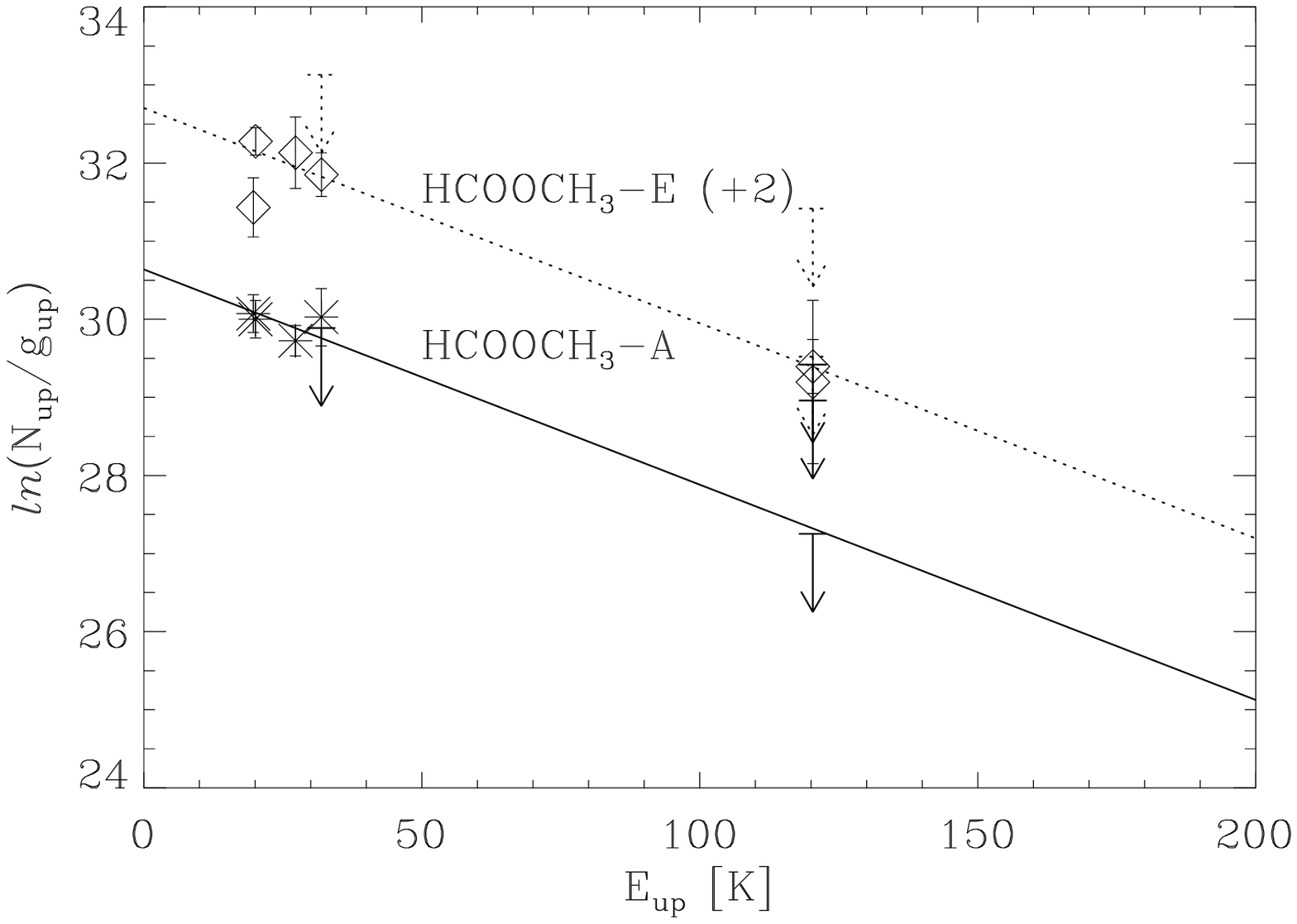}
\plotone{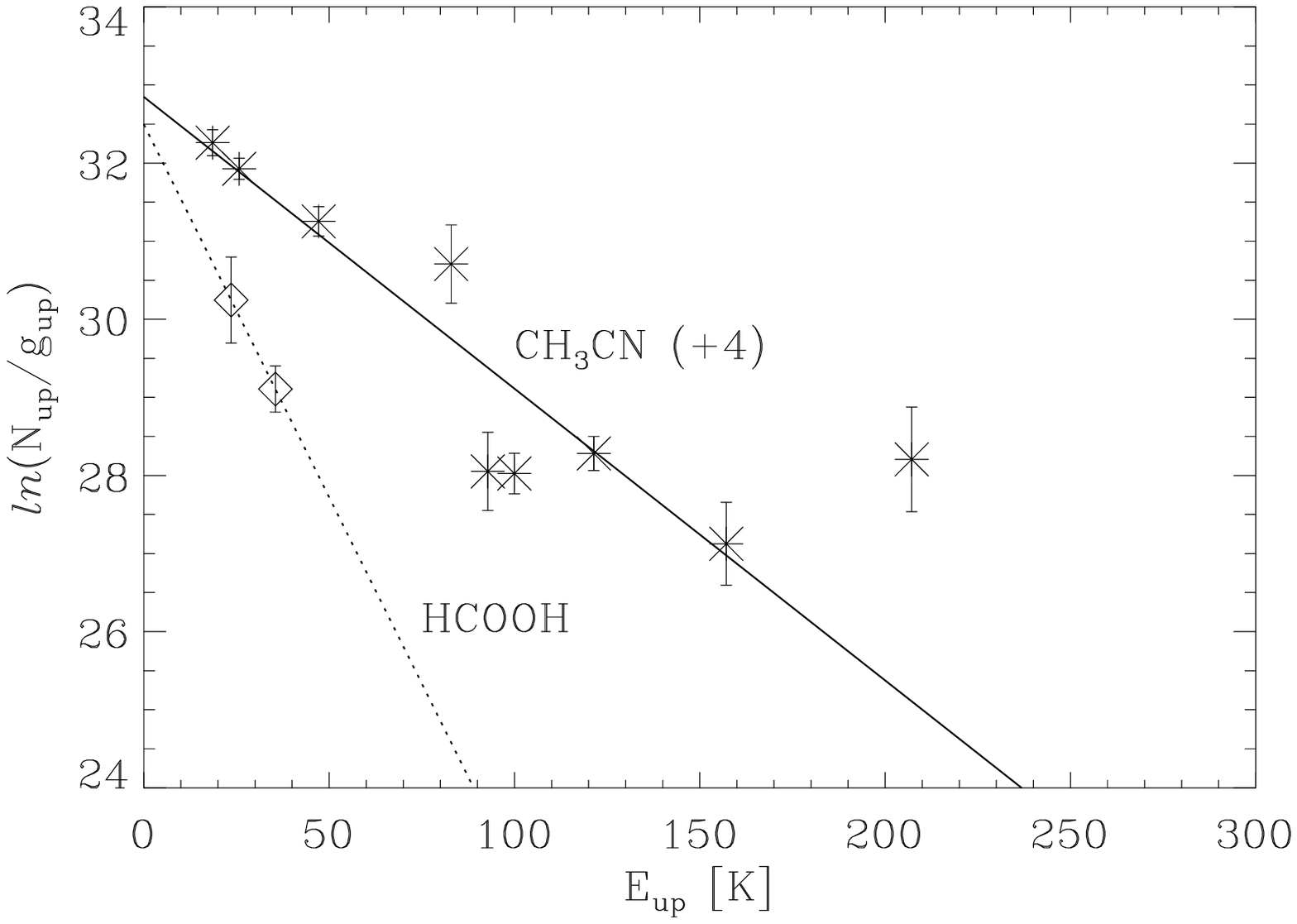}
\figcaption{Rotational diagrams of the detected molecules, corrected for 
beam dilution. 
The arrows show the upper limits for the transitions that have not 
been detected.
Lines represent the best fit to the data. Error bars are derived assuming a
calibration uncertainty of 10\% on top of the statistical error. 
The excess of emission of the CH$_3$CN transition at 210 K is probably 
due to contamination from unknown line(s). \label{rotdia}}
\end{figure}

Fig.~\ref{spectrum} shows two examples of low resolution spectra we obtained. 
Detected
transitions have been identified using the JPL molecular line catalog 
(Pickett et al. 1998) and are reported in Table~\ref{detections}.
We considered as good identifications only lines with a 3-$\sigma$ detection
and a $V_{\rm LSR}$=6.8$\pm$0.3 \kms. 
We detected three of the five targeted molecules: 10 transitions for 
HCOOCH$_3$ (A and E), 2 for HCOOH and 9 for CH$_3$CN. We also have a possible 
detection for C$_2$H$_5$OH at 90.118 GHz; unfortunately, no other transition
with a low enough energy and high enough Einstein coefficient was contained
within the frequency ranges we observed to confirm the correct identification. 
No transitions of CH$_3$OCH$_3$ and
C$_2$H$_5$CN were detected to a noise limit of 6 and 2 mK respectively.
All detected lines have linewidths $\sim 2-3$ \kms, with few exceptions,
likely due to the presence of unresolved triplets or to the contamination
of unidentified lines.
In order to derive the rotational temperature and column density, we built
rotational diagrams (Fig.~\ref{rotdia}) in 
which the observed fluxes were
corrected for beam dilution, assuming a source size of 0$\farcs$5 
(derived from 
a hot core radius of 53~AU and a distance of 220~pc, as found by 
Maret et al. 2004).
The assumption that the complex molecules are confined to
the hot corino is supported by a Plateau de Bure
interferometric study of IRAS16293 which shows localized emission
in a region $\sim1\farcs 4$
around the protostar (Bottinelli et al. in preparation).

The rotational temperatures, total column densities and abundances for the 
dectected molecules are presented in Table~\ref{T+N}. 
Note that the large errors in the HCOOH (this work) and CH$_3$OCH$_3$ 
(Cazaux et al. 2003) abundances are
due to a poor constraint of the rotational temperature, and hence 
column density, of these two molecules, even though each molecule is
clearly detected in each case.

\section{Discussion and conclusion}

The most important result of the present work is the detection of
complex molecules in the hot corino of IRAS4A, the second
Class 0 in which those molecules have been searched for, after IRAS16293
(Cazaux et al. 2003).  This result demonstrates that as soon as a warm
region is created in the center of the envelope of low mass protostars, 
complex molecules are readily formed and/or injected on timescales
lower than the estimated Class 0 source ages 
($\sim5\times10^4$~yr in IRAS16293 
and $\sim6500$~yr in IRAS4A; e.g. Maret et al. 2002), and, 
most importantly, shorter than
the transit time in the hot corinos. The latter is $\sim400$~yr and 
$\sim120$~yr in
IRAS16293 and IRAS4A respectively, based on the hot corino sizes 
quoted in Maret et al. (2004) and assuming free-falling gas.

We compare
the measured composition of the hot corino of IRAS4A to
IRAS16293 (Cazaux et al. 2003) and the massive hot core of
OMC-1 (Sutton et al. 1995) in Table~\ref{T+N}. 
Note that the latter abundances are derived from single dish measurements 
with a $14''$ beam, which encompasses several hot cores
(Wright et al. 1996). 
Unfortunately, not all the molecules considered here have interferometric
measurements available, so that we can only use
these $14''$ beam-averaged estimates of the abundances.

\begin{figure}
\epsscale{0.8}
\plotone{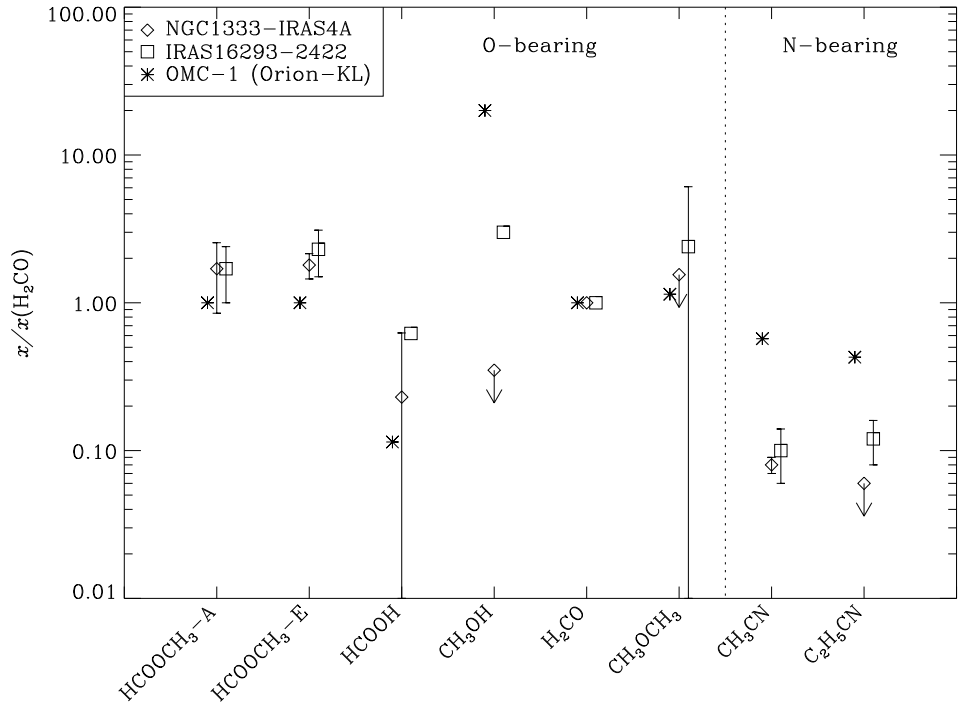}
\caption{The abundances of the observed species (reported on the x-axis)
normalized to the H$_2$CO abundances. 
Stars refer to the OMC-1 hot core, squares to the hot corino of IRAS16293 
and diamonds to the one of IRAS4A. Arrows represent upper limits in IRAS4A
derived from our observations. No errors were quoted by Cazaux et al. (2003)
for the HCOOH abundance, which was determined from two transitions only
and is rather uncertain.
\label{xratio}}
\end{figure}

The first remark is that the absolute abundances of the observed molecules
are one order of magnitude smaller in IRAS4A than in IRAS16293,
but their relative abundances with respect to H$_2$CO
are quite similar, with the exception of methanol,
which is underabundant with respect to H$_2$CO by about a factor 10 in
IRAS4A (Fig. \ref{xratio}).
There are two reasons to consider abundances with respect to
formaldehyde.
The first one is observational: while the IRAS16293 hot core has now been
imaged with the Plateau de Bure interferometer (Bottinelli et al.
in preparation) and its size confirmed to be $\sim1\farcs 4$, 
the IRAS4A core size is only indirectly
estimated from dust continuum single dish ($12''$) observations to be 
$0\farcs 5$
and no interferometric observations are available yet with such a high
resolution.
So the IRAS4A core size might be wrong by up to a factor three
(Maret et al. 2004) and the abundances by up to a factor ten, i.e.
the absolute abundances of IRAS4A could be comparable to those of IRAS16293.
Using abundance ratios allow us to remove this size uncertainty.
The second reason is theoretical: 
``standard'' hot core models predict that molecules like methyl formate
or methyl cyanide are second generation molecules formed in the warm gas
from the evaporated grain mantle constituents (formaldehyde, ammonia
and methanol: e.g. Charnley et al. 1992; Caselli et al. 1993;
Rodgers \& Charnley 2003). It is therefore interesting to compare the
abundances of the complex molecules to those of one of these supposed
parent molecules. Formaldehyde was chosen because we only have an upper
limit on the methanol abundance (Maret et al. in preparation) 
and no measurements of the ammonia abundance are available.

A possible interpretation for the similarity in the complex 
molecules' relative abundances, with respect to H$_2$CO and not 
with respect to CH$_3$OH,
is that the former is the 
mother molecule of the observed O-bearing species,
e.g. likely the case of HCOOCH$_3$ (Charnley et al. 1992), 
and that the chemical evolution
timescale is shorter than the age of the youngest source.
Charnley et al. (1992) also predict that methanol is
the mother molecule of CH$_3$OCH$_3$, but we cannot say whether
the available data confirm this hypothesis since we only have an upper limit
on the abundance of this molecule in IRAS4A and a large error in 
IRAS16293.
Similarly, the N-bearing molecules
CH$_3$CN and C$_2$H$_5$CN 
could both be daughters of the same mother molecule, probably ammonia.
This would imply that the two sources have a similar ammonia mantle abundance.
Alternatively, (some of ?) the reported
molecules are possible mantle constituents themselves.
This may be the case for formic acid, as predicted
by Tielens \& Hagen (1982), and suggested by the
observational study by Liu et al. (2001).
Moreover, the analysis of ISO absorption spectra towards the 
massive hot core W33A (e.g. Schutte et al. 1997)
is consistent with the presence of solid formic acid
and would also support the idea of this species being a mantle constituent.
However, these considerations do not take into account the evolutionary state
of the objects and the fundamental
question is: does the abundance of any of
these complex molecules have anything to do with the age
and/or evolutionary stage
of the protostar, or is it dominated by the initial mantle
composition?
Evidently, two sources are not enough to answer this question, and
observations of more low mass sources are required.

Regarding the comparison with the massive hot core(s) in Orion,
Fig. \ref{xratio} would suggest that, with respect to formaldehyde, 
there is a deficiency of methanol and of N-bearing complex molecules
in the low mass hot corinos.
It is possible that these differences are mostly due to a different
grain mantle composition, i.e. to a different pre-collapse density.
However, recall that the abundance ratios of CH$_3$CN and
CH$_3$OH in Figure \ref{xratio} refer to the $14''$ beam-averaged
values around the OMC-1 hot core, which in fact includes several
smaller cores (Wright et al. 1996). 
Therefore, in order to make precise comparisons, higher resolution
observations of the OMC-1 hot core are needed. It is also worth noting
that if we consider for example the measurements by Wright et al. 1996
in the Compact Ridge component (a region about $10''$ away from the
hot core central position, which is also a site of mantle evaporation
and of active gas phase chemistry; e.g. Charnley et al. 1992), the
CH$_3$CN and CH$_3$OH abundance ratios with respect to H$_2$CO are
(surprisingly) close to those found for the hot corinos of IRAS16293 and
IRAS4A. Hence, interferometric observations of a larger number of
massive hot cores are necessary to provide a significant comparison of
the hot corinos with their high mass counterparts.

In summary, although the present observations do not allow us to answer the
questions why and how complex molecules are formed,
they do show that hot corinos, in the wide definition
of chemically enriched regions, are a common property of solar-type
protostars in the early stages. 
The evidence is that the types of complex molecules that are formed
are determined primarily by the composition of the grain mantles.
At this stage, it is not clear
whether the evolutionary stage of the protostar plays any role at all,
other than governing the presence and size of the mantle evaporation
region.

\section{References}
\parskip=0pt
\bigskip

\pp Bacmann, A., Lefloch, B., Ceccarelli, C., Steinacker, J., Castets, A., 
\& Loinard, L. 2003, ApJ, 585, L55

\pp Blake, G. A., Sandell, G., van Dishoeck, E. F., Groesbeck, T. D., 
Mundy L. G., \& Aspin, C. 1995, ApJ, 441, 689

\pp Caselli, P., Hasegawa, T. I., \& Herbst, Eric 1993, ApJ, 408, 548

\pp Cazaux, S., Tielens, A. G. G. M., Ceccarelli, C., Castets, C. Wakelam, V.,
 Caux, E., Parise, B., \& Teyssier, D. 2003, ApJ, 593, L51

\pp Ceccarelli, C., Castets, A., Caux, E., Hollenbach, D., Loinard, L., 
Molinari, S., \& Tielens, A. G. G. M. 2000a, A\&A, 355, 1129

\pp Ceccarelli, C., Castets, A., Loinard, L., Caux, E., \& Tielens, A. G. G. M.
1998, A\&A, 338, L43

\pp Ceccarelli, C., Loinard, L., Castets, A., Tielens, A. G. G. M., Caux, E., 
Lefloch, B., \& Vastel, C. 2001, A\&A, 372, 998

\pp Ceccarelli, C., Loinard, L., Castets, A., Tielens, A. G. G. M., 
\& Caux, E. 2000b, A\&A, 357, L9

\pp \u{C}ernis, K. 1990, Ap\&SS, 166, 315

\pp Charnley, S. B. 1995, Ap\&SS, 224, 251

\pp Charnley, S. B., Rodgers, S. D. \& Ehrenfreund, P. 2001, A\&A, 378, 1024

\pp Charnley, S. B., Tielens, A. G. G. M., \& Millar, T. J. 1992, ApJ, 399, L71

\pp Choi, M., Panis, J.-F., \& Evans, N. J. 1999, ApJS, 122, 519

\pp Di Francesco, J., Myers, P. C., Wilner, D. J., Ohashi, N., \& 
Mardones, D. 2001, ApJ, 562, 770

\pp Lay, O. P., Carlstrom, J. E., \& Hills, R. E. 1995, ApJ, 452, L73

\pp Lefloch, B., Castets, A., Cernicharo, J., \& Loinard, L. 1998, 
ApJ, 504, L109

\pp Liu, S.-Y., Mehringer, D. M., \& Lewis, E. S. 2001, ApJ, 552, 654

\pp Maret, S., Ceccarelli, C., Caux, E., Tielens, A. G. G. M., J\o rgensen, 
J. K., van Dishoeck, E., Bacmann, A., Castets, A., Lefloch, B., Loinard, L., 
Parise, B., \& Sch\"oier, F. L. 2004, A\&A 416, 577

\pp Maret, S., Ceccarelli, C., Caux, E., Tielens, A. G. G. M. \&
Castets, A. 2002, A\&A, 395, 573
 
\pp Parise, B., Castets, A., Herbst, E., Caux, E., Ceccarelli, C., 
Mukhopadhyay, I., Tielens, A.G.G.M. 2004, A\&A 416, 159

\pp Parise, B., Ceccarelli, C., Tielens, A. G. G. M., Herbst, E., 
Lefloch, B., Caux, E., Castets, A., Mukhopadhyay, I., Pagani, L., \&
 Loinard, L. 2002, A\&A, 393, L49

\pp Pickett, H. M., Poynter, R. L., Cohen, E. A., Delitsky, M. L., 
Pearson, J. C., \& Muller, H. S. P., 1998
``Submillimeter, Millimeter, and Microwave Spectral Line Catalog'', 
J. Quant. Spectrosc. \& Rad. Transfer 60, 883

\pp Roberts, H., Fuller, G. A., Millar, T. J., Hatchell, J, \& Breckle, J. V.
2002, A\&A, 381, 1026

\pp Rodgers, S. D., \& Charnley, S. B. 2003, ApJ, 585, 355

\pp Rodgers, S. D., \& Charnley, S. B. 2001, ApJ, 546, 324

\pp Sandell, G., Aspin, C., Duncan, W. D., Russell, A. P. G., \& Robson, E. I.
1991, ApJ, 376, L17

\pp Sch\"oier, F. L., J\o rgensen, J. K., van Dischoeck, E. F., 
\& Blake, G. A. 2002, A\&A, 390, 1001

\pp Schutte, W. A., Greenberg, J. M., van Dishoeck, E. F., Tielens, 
A. G. G. M., Boogert, A. C. A., \& Whittet, D. C. B. 1997, Ap\&SS 255, 61

\pp Sutton, E. C. Peng, R., Danchi, W. C., Jaminet, P. A., Sandell, G., 
\& Russell, A. P. G. 1995, ApJ, 97, 455

\pp Tielens, A. G. G. M. \& Hagen, W. 1982, A\&A, 114, 245

\pp Vastel, C., Phillips, T. G., Ceccarelli, C., \& Pearson, J. 2003, 
ApJ, 593, L97

\pp Walmsley, C. M., Cesaroni, R., Churchwell, E., \& Hofner, P. 1992,
Astron. Gesellschaft Abstract Ser., 7, 93

\pp Wright, M. C. H., Plambeck, R. L., \& Wilner, D. J. 1996, ApJ, 469, 216

\clearpage

\clearpage

\clearpage

\clearpage

\begin{deluxetable}{lcrrccccr}
\tabletypesize{\scriptsize}
\tablecaption{Molecular lines detected toward IRAS4A. \label{detections}}
\tablehead{
\colhead{Molecule} & \colhead{Transition line} & \colhead{Frequency} &  \colhead{E$_u$}  & \colhead{$\delta V$\tablenotemark{a}} &
\colhead{$\Delta V$\tablenotemark{b}}   & \colhead{$T_{\rm mb}$}  & \colhead{$\int T_{\rm mb}$dV} & \colhead{rms}\tablenotemark{c} \\
\colhead{} &  \colhead{}  & \colhead{(MHz)}     & \colhead{(cm$^{-1}$)}  & \colhead{(km s$^{-1}$)} &
\colhead{(km s$^{-1}$)} & \colhead{(mK)}      & \colhead{(K km s$^{-1}$)}  & \colhead{(mK)}\\
}
\startdata
HCOOCH$_3$-A  & $7_{2,5}-6_{2,4}$                    &  90156.5  &  13.7  & 0.5 &  1.5 &  22  & 0.036   &  5 \\
              & $8_{0,8}-7_{0,7}$                    &  90229.7  &  13.9  & 0.5 &  2.5 &  16  & 0.041   &  5 \\
              & $8_{3,6}-7_{3,5}$                    &  98611.1  &  18.9  & 0.2 &  1.2 &  28  & 0.036   &  7 \\
              & $8_{4,5}-7_{4,4}$                    &  98682.8  &  22.2  & 1.9 &  4.2 &  95  & 0.042   &  2 \\
\tableline	    			                    	       		   		      	   	  	    
HCOOCH$_3$-E  & $7_{2,5}-6_{2,4}$                    &  90145.7  &  13.7  & 0.5 &  1.3 &  14  & 0.019   &  5 \\
              & $8_{0,8}-7_{0,7}$                    &  90227.8  &  14.0  & 0.5 &  3.1 &  16  & 0.055   &  5 \\
              & $8_{3,6}-7_{3,5}$                    &  98607.8  &  18.9  & 0.9 &  3.8 &  13  & 0.054   &  4 \\
              & $8_{4,5}-7_{4,4}$                    &  98711.7  &  22.2  & 0.2 &  1.4 &  22  & 0.034   &  7 \\
              & $20_{2,18}-19_{2,18}$                & 226713.1  &  83.6  & 0.8 &  1.0 &  89  & 0.099   & 26 \\
              & $20_{3,18}-19_{3,17}$                & 226773.3  &  83.6  & 0.8 &  2.1 &  54  & 0.121   & 19 \\
\tableline	    			                    	       		   		      	   	  	    
HCOOH         & $4_{2,2}-3_{2,1}$                    &  90164.5  &  16.4  & 0.5 &  0.8 &  16  & 0.015   &  5 \\
              & $6_{2,4}-5_{2,3}$                    & 135737.7  &  24.6  & 1.4 &  1.8 &  15  & 0.029   &  5 \\
\tableline	    			                     		     		   	      	      	      
CH$_3$CN\tablenotemark{d} & $6_{3,0}-5_{3,0}$ & 110364.6  &  57.6  & 0.8 &  5.1 &  20  & 0.110   &  6 \\
              & $6_{2,0}-5_{2,0}$ & 110375.1   &  32.8  & 0.8 &  2.3 &  46  & 0.112   &  6 \\
              & $6_{1,0}-5_{1,0}$ & 110381.5   &  17.9  & 0.8 &  3.4 &  67  & 0.241   &  6 \\
              & $6_{0,0}-5_{0,0}$ & 110383.6   &  12.9  & 0.8 &  4.3 &  76  & 0.347   &  6 \\
              & $14_{4,0}-13_{4,0}$& 257448.9  & 143.9  & 1.2 &  3.3 &  40  & 0.141   & 12 \\
              & $14_{3,0}-13_{3,0}$& 257482.7  & 109.1  & 0.4 &  2.6 &  53  & 0.150   & 19 \\
              & $14_{2,0}-13_{2,0}$& 257507.9  &  84.3  & 0.4 &  3.1 &  59  & 0.195   & 19 \\
              & $14_{1,0}-13_{1,0}$& 257522.5  &  69.4  & 0.4 &  2.2 &  74  & 0.172   & 19 \\
              & $14_{0,0}-13_{0,0}$& 257527.4  &  64.4  & 1.2 &  3.8 &  68  & 0.274   & 12 \\
\enddata
\tablenotetext{a}{Spectral resolution of the observation 
(when possible, the integrated intensity was derived from the high
resolution data).}
\tablenotetext{b}{Width of the observed line.}
\tablenotetext{c}{rms computed over the linewidth.}
\tablenotetext{d}{All the CH$_3$CN lines are (unresolved) triplets. The 
quoted signal is the integral over each triplet. Larger linewidths
could be due to the larger spacing between the components of the triplets.}

\end{deluxetable}

\clearpage

\begin{deluxetable}{lcccccc}
\tabletypesize{\scriptsize}
\tablewidth{0pt}
\tablecaption{Results from the rotational diagrams for IRAS4A, in 
comparison with IRAS16293 and the massive hot core OMC-1. \label{T+N}}
\tablehead{
\colhead{} & \multicolumn{3}{c}{IRAS4A} & \colhead{} & \colhead{IRAS16293\tablenotemark{b}}  & \colhead{OMC-1\tablenotemark{c}} \\
\cline{2-4}\cline{6-7}

\colhead{Molecule} & \colhead{$T_{\rm rot}$} & \colhead{$N_{\rm total}$} & \colhead{$x$\tablenotemark{a}} & \colhead{}& 
\multicolumn{2}{c}{$x$}\\
\colhead{}         &  \colhead{    (K)     } & \colhead{   (cm$^{-2}$) } &  \colhead{}    & &   \colhead{}      & \colhead{}
}
\startdata
HCOOCH$_3$-A  & 36\tablenotemark{d} & 5.5 $\pm$ 2.7(16) & 3.4 $\pm$ 1.7(--8) & & 1.7 $\pm$ 0.7 (--7) & 1(--8)\\
HCOOCH$_3$-E  & 36 $\pm$ 5          & 5.8 $\pm$ 1.1(16) & 3.6 $\pm$ 0.7(--8) & & 2.3 $\pm$ 0.8 (--7) & 1(--8)\\
HCOOH         & 10 $\pm$ 6          & 7.3 $\pm$13.0(15) & 4.6 $\pm$ 7.9(--9) & & $\sim$6.2(--8)      & 8(--10)\\
CH$_3$CN      & 27 $\pm$ 1          & 2.6 $\pm$ 0.3(15) & 1.6 $\pm$ 0.2(--9) & & 1.0 $\pm$ 0.4 (--8) & 4(--9)\\

CH$_3$OH      & ...                 & ... & $\leq$7(--9)\tablenotemark{f} & & 3(--7) & 1(--7)\\
H$_2$CO       & ...                 & ... &       2(--8)\tablenotemark{g} & & 1(--7)\tablenotemark{h} & 7(--9)\\
\tableline
\multicolumn{7}{c}{Upper Limits}\\
\tableline
CH$_3$OCH$_3$ & 36\tablenotemark{d} & $\leq$4.5(16) & $\leq$2.8(--8) & & 2.4 $\pm$ 3.7(--7) & 8(--9)\\
C$_2$H$_5$CN  & 27\tablenotemark{e} & $\leq$1.9(15) & $\leq$1.2(--9) & & 1.2 $\pm$ 0.4(--8) & 3(--9)\\
\enddata

\tablenotetext{a}{Assuming an H$_2$ column density in the hot corino of $N$(H$_2$) = 1.6 $\times$ 10$^{24}$ cm$^{-2}$
(From Maret et al. 2004).}
\tablenotetext{b}{From Cazaux et al. 2003.}
\tablenotetext{c}{From Sutton et al. 1995.}
\tablenotetext{d}{$T_{\rm rot}$ assumed to be similar to the one derived for HCOOCH$_3$-E.}
\tablenotetext{e}{$T_{\rm rot}$ assumed to be similar to the one derived for CH$_3$CN.}
\tablenotetext{f}{From Maret et al. 2004, in prep.}
\tablenotetext{g}{From Maret et al. 2004.}
\tablenotetext{h}{From Ceccarelli et al. 2000b.}
\end{deluxetable}

\end{document}